\documentclass[twocolumn,superscriptaddress,showpacs,preprintnumbers,amsmath,nofootinbib,amssymb]{revtex4}
\usepackage{graphicx}
\usepackage{dcolumn}
\usepackage{amssymb}
\usepackage{amsmath}
\usepackage{amsfonts}
\usepackage{amsbsy}
\usepackage{color}
\usepackage{rotating}
\usepackage{epsfig}
\usepackage[english]{babel}

\renewcommand{\[}{\begin{equation}}
\renewcommand{\]}{\end{equation}}

\begin{document}

\title{The Inflationary Origin of the Cold Spot Anomaly}

\author{Juan C. Bueno S\'anchez}
%\email{jcbueno@fis.ucm.es}
\email{juan.c.bueno@correounivalle.edu.co}
\affiliation{Departamento de F\'isica, Universidad del Valle, A.A. 25360, Santiago de Cali, Colombia.}
\affiliation{Centro de Investigaciones en Ciencias  B\'asicas y Aplicadas, Universidad Antonio Nari\~no, Cra 3 Este \# 47A-15, Bogot\'a D.C. 110231, Colombia.}
\affiliation{Escuela de F\'isica, Universidad Industrial de Santander, Ciudad Universitaria, Bucaramanga 680002, Colombia.}
\affiliation{Departamento de F\'isica At\'omica, Molecular y Nuclear,
Universidad Complutense de Madrid, 28040, Madrid, Spain.}

\begin{abstract}
Single-field inflation, arguably the simplest and most compelling paradigm for the origin of our Universe, is strongly supported by the recent results of the Planck satellite and the BICEP2 experiment. The results from Planck, however, also confirm the presence of a number of anomalies in the Cosmic Microwave Background (CMB), whose origin becomes problematic in single-field inflation. Among the most prominent and well-tested of these anomalies is the Cold Spot, which constitutes the only significant deviation from Gaussianity in the CMB. Planck's non-detection of primordial non-Gaussianity on smaller scales thus suggests the existence of a physical mechanism whereby significant non-Gaussianity is generated on large angular scales only. In this letter, we address this question by developing a \emph{localized} version of the inhomogeneous reheating scenario, which postulates the existence of a scalar field able to modify the decay of the inflaton on localized spatial regions only. We demonstrate that if the Cold Spot is due to an overdensity in the last scattering surface, the localization mechanism offers a feasible explanation for it, thus providing a physical mechanism for the generation of \emph{localized} non-Gaussianity in the CMB. If, on the contrary, the Cold Spot is caused by a newly discovered supervoid (as recently claimed), we argue that the localization mechanism, while managing to enhance underdensities, may well shed light on the rarity of the discovered supervoid.

\end{abstract}


%\pacs{$98.80.-k,11.25.Uv,98.80.Cq,02.40.Xx$}

%--------------------------------------------------------------------------------
%--------------------------------------------------------------------------------

\maketitle

\section{Introduction}

After the recent results from Planck \cite{Ade:2013zuv,Ade:2013uln,Ade:2013nlj,Ade:2013ydc} and BICEP2 \cite{Ade:2014xna}, the inflationary paradigm, and in particular single-field inflation, seems to be the one chosen by nature to generate the observed adiabatic, nearly scale-invariant, gaussian spectrum of curvature perturbations and the B-mode polarization at degree angular scales. Nevertheless, a number of large-angle anomalies have been confirmed by Planck \cite{Ade:2013nlj}, which seems to pose a relative challenge for single-field inflation.

In this letter we pay particular attention to the Cold Spot anomaly; a large, nearly circular region of the CMB sky, around $10^\circ$ in angular size in the southern hemisphere, with a significant temperature decrement (see \cite{Vielva:2010ng} for an extensive review). This anomaly was first detected in 2004 \cite{Vielva:2003et}, and since then it has been the subject of numerous statistical analysis \cite{G11,morpho,Vielva:2007kt}. Similarly to the anomalies of the low quadrupole and the alignment of the low multipoles \cite{Francis:2009pt}, the Cold Spot has been argued to be of no statistical significance \cite{Zhang:2009qg}. However, an intriguing observation first put forward in \cite{Vielva:2007kt} is that there seems to exist an anomalous hot spot in the CMB too. In fact, the results from Planck confirm the existence of several other anomalous hot and cold spots \cite{Ade:2013nlj}. Although detected to a smaller significance, their presence persists after applying different masks to the data \cite{Ade:2013nlj}. This indicates that, provided the anomalous nature of the Cold Spot is confirmed \cite{Vielva:2010vn}, one should envisage a mechanism flexible enough to accommodate a number of anomalous hot and cold spots. On the theoretical front, on the other hand, a number of alternatives have been considered in the literature. Briefly, the most significant consider the Cold Spot as the result of: a local void \cite{voids}, the Sunyaev-Zeldovich effect \cite{texture2}, the formation of a cosmic texture \cite{texture1}, multifield inflation \cite{Afshordi:2010wn}, or chaotic preheating \cite{Bond:2009xx}. At the time of writing, two simultaneous papers appeared that went unnoticed to the author\footnote{I thank R. Brandenberger for pointing these out.} \cite{Finelli:2014yha,Szapudi:2014zha}. In these, the detection (with 5$\sigma$-6$\sigma$ significance) of a supervoid aligned with the Cold Spot is reported and investigated as the origin of the Cold Spot itself via a Rees-Sciama effect. Although the discovered supervoid constitutes a plausible explanation of the Cold Spot, its extreme size, around $200h^{-1}\,$Mpc in radius (constituting a $3.5 - 5\sigma$ fluctuation of the $\Lambda$CDM model) demands that this is confirmed by further studies. In this sense, although the LTB fit carried out in \cite{Finelli:2014yha} (see also \cite{Szapudi:2014eza}) to the discovered supervoid is claimed to provide a perfect explanation of the Cold Spot, such a fit, in itself,  does not address the rarity of the supervoid, which then remains an open question. As we argue later on, the mechanism here described might shed light on the rarity of such an extreme void. In any case, a number of other anomalous spots of similar size have been reported by Planck, and hence it is of fundamental importance to investigate the extent to which inflationary fluctuations can give rise to such spots in the CMB.

Since observations clearly support an adiabatic, nearly scale-invariant, gaussian spectrum of curvature perturbations (according to the generic predictions of single-field inflation), in this letter we describe in detail a physical mechanism for the generation of the Cold Spot inspired by the idea that the latter, being the most prominent non-gaussian signal in the CMB, has its origin in an isocurvature fluctuation. Hence, we take the view that the curvature perturbation imprinted by the inflaton is supplemented with an additional contribution providing the Cold Spot signal. The mechanism developed here is based on the spatial modulation obtained by an interacting light field during inflation. In our setting, the spatial modulation arises as the result of a trapping mechanism experienced by the light field due to its coupling to other degrees of freedom. Since light fields undergo inflationary fluctuations, the trapping does not occur everywhere at the same time during inflation. Owing to this, it is possible that the field relaxes to its equilibrium value (due to its interactions with other fields) in some locations, whereas in others, the field manages to evade its trapping and retains an initially large expectation value, which we justify conveniently. To convert the modulation in the isocurvature field into a curvature perturbation we utilize the inhomogeneous reheating scenario, thus assuming that the isocurvature field controls the decay rate of the inflaton. In contrast to the usual inhomogeneous reheating, in our scenario, owing to the modulation obtained by the isocurvature field, the contribution to the curvature perturbation is imprinted on \emph{localized} regions of the CMB only. We show how this \emph{localization} mechanism allows us to account for the Cold Spot while respecting the stringent bounds on non-Gaussianity imposed by Planck.

The letter is organized as follows. In Sec.~\ref{sec2}, we describe the main idea and basic working of the mechanism. In Sec.~\ref{sec3}, we study the stochastic behavior of the field responsible for the emergence of the Cold Spot and quantify the number density of anomalous spots in the CMB. In Sec.~\ref{sec4}, we estimate the curvature perturbation contributed by our \emph{localized} version of the inhomogeneous reheating and constrain the model parameters accordingly. Conclusions to this letter are presented in Sec.~\ref{sec5}.

\section{A mechanism for the Cold Spot}\label{sec2}
We investigate a system of two interacting, massive scalar fields, $\sigma$ and $\chi$, minimally coupled to gravity and whose energy density remains always subdominant. Taking an interaction term of the form $g^2\sigma^2\chi^2$ and setting aside the interactions of $\sigma$ and $\chi$ with other fields, the Lagrangian of the system is
\[\label{eq27}
{\cal L}=\frac12\partial_\mu\sigma\partial^\mu\sigma+\frac12\partial_\mu\chi\partial^\mu\chi
-\frac12 \bar m_\sigma^2\sigma^2-\frac12 \bar m_\chi^2\chi^2-\frac12\,g^2\sigma^2\chi^2,
\]
where $g$ is a coupling constant. The above interaction $g^2\sigma^2\chi^2$ is ubiquitous in quantum field theory, and its consequences have been extensively studied in the theory of reheating and preheating \cite{G9}. Moreover, this coupling results in a trapping mechanism whereby points of enhanced symmetry become a preferred location (in field space) for string moduli \cite{G1}. The trapping mechanism has been employed in inflation model building (trapped inflation) \cite{G7}, to generate non-Gaussianity of the inflaton's perturbation spectrum \cite{G8}, and more recently to study the stochastic evolution of coupled flat directions \cite{Sanchez:2012tk}. In our setting, we take advantage of a result pointed out in \cite{Sanchez:2012tk}, namely, that the trapping mechanism gives rise to a spatial modulation in one of the fields involved in the coupling, provided it begins with an expectation value large enough to make the other field heavy.

\subsection{The spatial modulation of $\sigma$}\label{sec2a}
As shown in \cite{Sanchez:2012tk}, when two scalar fields are subject to a coupling of the form $g^2\sigma^2\chi^2$ in an inflationary background, one of the fields ($\sigma$ in the following) manages to fluctuate as a free field provided its expectation value is sufficiently large for the other field ($\chi$) to become heavy. In \cite{Sanchez:2012tk} it was assumed that the mass of $\chi$ is mainly determined by its interaction with $\sigma$, hence the condition for $\sigma$ to fluctuate as a free field is $m_\chi^2 \simeq g^2\sigma^2\gg H^2$. We then assume that at $t=t_*$, when the scale of the observable Universe exits the horizon $N_*$ $e$-foldings before the end of inflation, the field's expectation value $\sigma_*$ satisfies this requirement. Since $\sigma$ is light enough to undergo particle production, as the last phase of slow-roll progresses, the field fluctuates similarly to a free field, growing larger in some locations and smaller in others. The trapping mechanism is triggered after $\sigma$, owing to both its dynamics and random fluctuations, decreases enough for the $\chi$ field to be produced during inflation. This happens when $\sigma\sim \sigma_c$, where we introduce the crossover scale \cite{Sanchez:2012tk}
\[\label{eq26}
\sigma_c\equiv \sqrt{10}\,g^{-1}H\,.
\]
This mechanism confines $\sigma$ to the origin of its potential, where it fluctuates indefinitely with an expectation value typically far smaller than before its trapping \cite{G1}.

Owing to the inflationary fluctuations, the trapping of $\sigma$ does not happen everywhere at the same time. Therefore, as slow-roll inflation progresses, $\sigma$ obtains a spatial modulation which can be schematically described as: $\sigma=\sigma_{\rm out}\gtrsim\sigma_c$ (where ``out'' stands for out-of-equilibrium or outstanding) in regions where $\sigma$ fluctuates similarly to a free field, and $\sigma=\sigma_{\rm eq}<\sigma_c$ (where ``eq'' stands for equilibrium) in the remaining regions where $\sigma$ is trapped due to its interaction with $\chi$. Regarding the magnitude of $\sigma_{\rm eq}$, it is important to note that the strength of the trapping mechanism is enhanced by the multiplicity of the $\chi$ field. For example, if $\chi$ belongs to a large GUT group one can expect $\sigma_{\rm eq}\ll\sigma_c$ shortly after the trapping mechanism is set off. In any case, if $\sigma$ becomes heavy after the trapping, and in the following we assume this is indeed the case, its expectation value becomes exponentially suppressed as inflation progresses\footnote{The magnitude of this suppression is studied in detail in \cite{BS}.}. Therefore, at the end of inflation $\sigma$ features a spatial modulation (obtained during the last $N_*$ $e$-foldings) such that
\[\label{eq9}
\sigma_{\rm eq}\ll\sigma_{\rm out}\,.
\]

The details of the modulation are investigated in the next section, but for now we discuss the large value $\sigma_*$ required for the modulation to arise, i.e. \mbox{$g\sigma_*\gg H_*$}. In principle, such condition is not problematic if $m_\sigma\ll H$, for in that case the equilibrium value of $\sigma$ in de Sitter space is anomalously large \cite{Starobinsky:1994bd}. However, in the context of the inhomogeneous reheating scenario, an important remark concerning $m_\sigma$ is in order. On the one hand, the recent discovery of the primordial B-mode polarization by the BICEP2 experiment, setting the tensor-to-scalar ratio to $r\simeq0.20$ \cite{Ade:2014xna}, implies that the inflaton excursion during inflation is above the Planck scale \cite{Lyth:1996im}. If $\sigma$ couples to the inflaton (to modulate its decay rate) through renormalizable interactions, then $\sigma$ receives a large correction to $m_\sigma^2$. On the other hand, if $\sigma$ couples to the inflaton through non-renormalizable interactions, then $m_\sigma^2\sim H^2$ as long as the time-averaged vacuum energy is positive \cite{Dine:1995uk}. Therefore, in the following we consider that $m_\sigma^2$ is dominated by Hubble-induced corrections, namely
\[\label{eq13}
m_\sigma^2\simeq c_\sigma H^2\,,
\]
where\footnote{This mass limit has also been studied in relation to large non-Gaussianities in quasi-single field inflation in \cite{Chen:2009we}.} $c_\sigma\sim{\cal O}(1)$. As a result, $\sigma$ evolves under the time-dependent quadratic potential $V(\sigma)\simeq \frac12\,c_\sigma H^2\sigma^2$. The lack of tuning entailed by allowing $c_\sigma\sim{\cal O}(1)$, although expected on theoretical grounds, immediately raises concerns as to the naturalness of the value $\sigma_*$ necessary to achieve the modulation in Eq.~(\ref{eq9}). The reason is that fields with masses on the Hubble scale typically have expectation values on the same scale (see Eq.~(\ref{eq23})), which is incompatible with the condition $g\sigma_*\gg H_*$. However, as shown below, this difficulty can be successfully addressed under certain, reasonable assumptions.

\subsection{Generating the initial condition}\label{sec2b}
Among the natural assumptions on the beginning of inflation is that it kicks off at energies close to the Planck scale in some sort of non-slow-roll phase. The initial phase, which occurs with the observable Universe within the horizon, is called \textit{primary} inflation \cite{Lyth:1998xn}. Usually, primary inflation is thought to set the initial conditions for the subsequent phase of slow-roll inflation. Such is the case of fast-roll inflation \cite{Linde:2001ae}, for example. The initial phase, however, has been argued to be of little interest (compared to the last phase of slow-roll) since the scales that exited the horizon during that epoch are well outside the present horizon. Nevertheless, despite this judgement, we investigate the conditions under which a non-slow-roll phase of primary inflation sets the appropriate initial conditions for the emergence of the Cold Spot. This approach thus suggests the use of this well-tested anomaly as a tool to constrain the primary epoch of inflation.

To illustrate this approach we consider a non-slow-roll phase characterized by a constant $\epsilon\equiv-\dot H/H^2\lesssim {\cal O}(1)$ for simplicity. This phase can be motivated, for example, by an inflaton with a non-negligible kinetic density due to its evolution under a steep potential\footnote{Steep scalar potentials are known to be ubiquitous in string theory. In turn, the results from BICEP2 \cite{Ade:2014xna} on the tensor-to-scalar ratio, implying an inflaton field in the Planck scale during inflation \cite{Lyth:1996im}, may be understood as an indication to consider string models of inflation (see \cite{Baumann:2014nda} for a recent review). In that case, a non-slow-roll phase of primary inflation can be easily motivated.}. Denoting by $H_0$ the scale at the beginning of inflation, the evolution of the background geometry is given by
\[\label{eq11}
H=H_0a(t)^{-\epsilon}\,\,,\,\,a(t)=(1+\epsilon H_0t)^{1/\epsilon}\,.
\]
We further assume that $\sigma$ begins the non-slow-roll phase with a vanishing expectation value, $\sigma_0=0$. Later non-vanishing values of $\sigma$ thus arise due to the accumulation of superhorizon modes. Using Eq.~(\ref{eq11}), the mode equation
\[
\ddot{\delta\sigma_k}+3H\dot{\delta\sigma_k}+\left(\frac{k^2}{a^2}+m_\sigma^2\right)\delta\sigma_k=0
\]
can be solved exactly. Imposing the flat spacetime vacuum solution in the subhorizon limit $k/aH\to\infty$ we find
\[\label{eq6}
\delta\sigma_k(t)=a^{-1}e^{\frac{i\pi}2\left(\nu+\frac12\right)}
\sqrt{-\frac{\pi\tau}{4}}\,H_\nu^{(1)}\left(-k\tau\right)\,,
\]
where $\tau=-[(1-\epsilon)aH]^{-1}$ is the conformal time and
\mbox{$\nu^2\equiv\frac94-\frac{c_\sigma-\epsilon(3-2\epsilon)}{(1-\epsilon)^2}$}. In the superhorizon regime, \mbox{$k/aH\to0$}, the mode $\delta\sigma_k$ scales as
\[
\delta\sigma_k\propto a^{-\alpha}\quad,\quad \alpha\equiv\frac32-\nu+\epsilon\left(\nu-\frac12\right)\,.
\]
Since $H\propto a^{-\epsilon}$, the ratio $\delta\sigma_k/H\propto a^{\epsilon-\alpha}$ grows exponentially during inflation when $\epsilon-\alpha>0$. The ensuing growth of $\sigma$ can be understood similarly to a tachyonic instability due to the shape of the scalar potential. In our case, however, the instability arises as a result of the rapid evolution of the background geometry.

Using Eq.~(\ref{eq6}), we compute the perturbation spectrum
\[\label{eq29}
{\cal P}_{\delta\sigma}(k)\equiv\lim_{k/aH\to0}\frac{k^3|\delta\sigma_k|^2}{2\pi^2}=\gamma\,\frac{H^2}{4\pi^2}\left(\frac{k}{aH}\right)^{3-2\nu}\,,
\]
where $\gamma=\frac{2^{-1+2\nu}\Gamma(\nu)^2}{\pi(1-\epsilon)^{1-2\nu}}$, and the variance\footnote{A result similar to Eq.~(\ref{eq12}) can be obtained using the stochastic approach to inflation applied to $\sigma$ \cite{BS}.}
\[\label{eq12}
\Sigma^2\equiv\langle(\sigma-\bar\sigma)^2\rangle=\gamma\frac{H^2}{4\pi^2(3-2\nu)}\left(1-e^{-(3-2\nu)N}\right)\,,
\]
where $N$ is the number of elapsed $e$-foldings since the beginning of inflation. In the above we used $\bar\sigma=0$, which follows from $\sigma_0=0$.

Coming back to the issue on the naturalness of $\sigma_*$, we may consider a particular initial condition $\sigma_*$ justified provided $\Sigma^2\gtrsim\sigma_*^2$ by the end of the non-slow-roll phase. From Eq.~(\ref{eq12}), this happens when $2\nu>3$ and $N$ is sufficiently large. In such case, using Eq.~(\ref{eq11}) and the minimum $N$ required to fulfill $\Sigma^2\gtrsim\sigma_*^2$, we may compute the minimum $H_0$ compatible with the generation of $\sigma_*$. To be precise, let us assume that $\Sigma^2=\sigma_*^2$ is fulfilled right at the onset of the slow-roll phase, when the observable Universe exits the horizon at $H_*\simeq2\times10^{14}\,$GeV. Taking for example $\epsilon=0.30$, $c_\sigma=0.15$, $g\sigma_*\simeq60H_*$ and $g$ in the range $0.1\leq g\leq 1$, we find the corresponding ranges $23\geq N\geq 17$ and $2.4\times10^{17}\,{\rm GeV}\geq H_0\geq 3.4\times10^{16}\,{\rm GeV}$, thus setting the beginning of inflation close to the Planck scale, as suggested in \cite{Lyth:1998xn}. Of course, the ratio $\sigma_*/H_*$ cannot be arbitrarily large. If we require the field $\sigma$ to be subdominant by the onset of slow-roll inflation, then $\rho_\sigma\ll H_*^2m_P^2$. Neglecting the kinetic density for simplicity, we obtain \mbox{$\sigma_*/H_*\ll c_\sigma^{-1/2}(m_P/H_*)\sim10^4c_\sigma^{-1/2}$}, thus validating the value of $\sigma_*$ previously chosen.

\subsection{\emph{Localized} inhomogeneous reheating and the generation of the Cold Spot}\label{sec2c}
To generate the Cold Spot we need a mechanism to convert the spatial modulation in $\sigma$ into a curvature perturbation. A simple option to achieve this is through the inhomogeneous reheating hypothesis \cite{G2}, which considers that a field undergoing inflationary fluctuations ($\sigma$ in our case) determines the inflaton decay rate, denoted by $\Gamma(\sigma)$. In this scenario, the contribution to the curvature perturbation on uniform density slices is given by \cite{G2,G3}
\[\label{eq28}
\zeta_\sigma=\alpha\left.\frac{\delta \Gamma(\sigma)}{\Gamma(\sigma)}\right|_{\rm dec}\,,
\]
where ``dec'' indicates the time of inflaton decay and $\alpha$ is the efficiency parameter. In the following, we consider that inflation gives way to a Universe dominated by the matter-like oscillations of the inflaton. Also, we assume that the inflaton decays much after inflation, which corresponds to $\alpha\simeq1/6$ \cite{G2,G3}.

To illustrate our \emph{localized} version of the inhomogeneous reheating, let us consider the decay rate \cite{Postma:2003jd}
\[\label{eq2}
\Gamma(\sigma)=\Gamma_0\left[1+\left(\frac{\sigma}{M}\right)^q\right]^r\,,
\]
where $\Gamma_0$ is the unperturbed inflaton's decay rate, $q\geq1$, $M$ is a mass scale, which we treat as a free parameter, and $\sigma< M$ at the time of decay. The central idea behind the \emph{localized} mechanism is that $\sigma_{\rm out}$ (corresponding to an outstanding value of $\sigma$) is large enough so that the corresponding $\zeta_\sigma$, given by
\[\label{eq20}
\zeta_\sigma\simeq\alpha q r\left(\frac{\sigma}{M}\right)^q\left.\frac{\delta\sigma}{\sigma}\,\right|_{\rm dec}\,,
\]
represents a sizable contribution to the total curvature perturbation, whereas $\sigma_{\rm eq}$ (with $\sigma_{\rm eq}\ll\sigma_{\rm out}$ as discussed in Sec.~\ref{sec2a}) results in a negligible contribution. At this point, it is important to remark that out-of-equilibrium patches may either persist until the time of inflaton decay (implying that Eq.~(\ref{eq9}) holds), or disappear, if $\sigma$ undergoes a non-perturbative decay after inflation, for example. Nevertheless, since a number of alternatives exist to prevent the non-perturbative decay of $\sigma$, in the following we assume that out-of-equilibrium patches do survive until reheating. In that case, it follows from Eq.~(\ref{eq20}) that the curvature perturbation $\zeta_\sigma$ inherits the spatial modulation of $\sigma$, namely
\[\label{eq7}
\zeta_{\rm eq}\ll \zeta_{\rm out}\,.
\]
In view of Eq.~(\ref{eq20}), one might think that for $q=1$, when $\zeta_\sigma$ depends on the perturbation $\delta\sigma$ only, the mechanism becomes inoperative since the spatial modulation of $\sigma$ is not transferred to $\zeta_\sigma$. However, since the potential for $\sigma$ is quadratic, the ratio $\delta\sigma/\sigma$ remains constant, and consequently $\zeta_\sigma$ becomes modulated as in Eq.~(\ref{eq7}) since $(\delta\sigma)_{\rm eq}\ll (\delta\sigma)_{\rm out}$.

The growth of the decay rate in Eq.~(\ref{eq2}) with $\sigma$ implies an anticipated decay in out-of-equilibrium regions, where $\sigma$ features an outstanding value. As a result, the energy density in these regions experiences an enhanced redshift compared to equilibrium ones, where the smallness of $\sigma_{\rm eq}\ll\sigma_{\rm out}$ causes a negligible perturbation of the inflaton decay. Consequently, out-of-equilibrium patches result in enhanced underdensities, which appear randomly in the observable Universe with the number density in Eq.~(\ref{eq4}). It is thus feasible that one such enhanced underdensity becomes the seed of the Cold Spot if it results in the formation of the discovered supervoid of about $200h^{-1}\,$Mpc in radius at redshift $z\simeq0.2$ \cite{Finelli:2014yha,Szapudi:2014zha,Szapudi:2014eza}. On the other hand, when enhanced underdensities intersect the last scattering surface, anomalous hot spots are generated in the CMB. Remarkably, an anomalous hot spot was identified in the WMAP data \cite{Vielva:2007kt}, whereas the more recent analysis of the Planck data suggests the existence of two anomalous hot spots \cite{Ade:2013nlj}. Therefore, from the qualitative point of view, the localized inhomogeneous reheating, along with the decay in Eq.~(\ref{eq2}), might suffice to imprint the pattern of anomalous spots observed in the CMB through enhanced underdensities along the line of sight (cold spots) and in the last scattering surface (hot spots) in just one strike.

In this letter, however, we focus on the possibility that the Cold Spot is caused by an enhanced overdensity in the last scattering surface, for which it is necessary to consider a decay rate different from that in Eq.~(\ref{eq2}). For example, if the inflaton undergoes a 2-body decay into $\psi$ particles, the corresponding rate is \cite{Postma:2003jd}
\[\label{eq14}
\Gamma=\Gamma_0\left[1-\left(\frac{2m_\psi}{m_\phi}\right)^2\right]^{1/2}\,,
\]
where $m_\psi=\lambda\sigma$ and $\lambda$ is a dimensionless coupling. Since $\Gamma$ decreases as $\sigma$ grows large, out-of-equilibrium patches undergo a suppressed redshift due to the delayed inflaton decay, thus resulting in enhanced overdensities. The contribution to the curvature perturbation in this case can be obtained from Eq.~(\ref{eq20}) after taking $q=2$, $r=1$ and performing the substitution $M\to\sqrt{qr}\lambda^{-1}m_\phi$.

From the above discussion, it follows that the checklist to account for the Cold Spot through an enhanced overdensity in the last scattering surface encompasses the following requirements. First, as we just discussed, the inflaton decay rate must have the appropriate dependence with $\sigma$ to actually generate a Cold Spot. Secondly, large expectation values $\sigma_{\rm out}$  must be correlated on scales comparable to the Cold Spot. At this point, we should remark that, in principle, field correlations with $\sigma\gtrsim\sigma_c$ may appear on scales other than the corresponding to the Cold Spot. Therefore, one might expect to find other anomalous spots in the CMB within a range of angular sizes. We address this important observation in Sec.~\ref{sec3}, where we compute the probabilistic distribution of out-of-equilibrium regions that appear in the CMB. The third requirement is that $\zeta_\sigma$ is sufficiently large in out-of-equilibrium regions to affect the curvature perturbation imprinted by the inflaton field, i.e. $\zeta_\sigma\sim4.8\times10^{-5}$ \cite{Ade:2013zuv,Ade:2013uln}. The constraints on the model parameters following from this requirement are obtained in Sec.~\ref{sec4}, after discussing the post-inflation evolution of $\sigma$.

\section{Stochastic description of the spatial modulation}\label{sec3}
As previously discussed, in order to account for the Cold Spot through localized inhomogeneous reheating, it is first necessary that an out-of-equilibrium patch of the appropriate size arises in the CMB. The goal of this section is to estimate the number density of spatial patches in which outstanding values $\sigma\gtrsim\sigma_c$ are correlated on a given comoving scale $k^{-1}$. For the sake of brevity, in the following we refer to the latter as \emph{$k$-patches}.

We assume that the scale corresponding to the ob\-ser\-vable Universe ($k={\cal H}_*$) exits the horizon $N_*$ $e$-foldings before the end of inflation, and that the field's expectation value in the Hubble-sized patch from which our observable Universe emerges is $\sigma_*>\sigma_c$. To obtain the distribution of $k$-patches at the end of inflation we need to describe the probabilistic evolution of $\sigma$ during slow-roll, which is dictated by the Fokker-Planck equation \cite{Starobinsky:1986fx}
\[\label{eq15}
\frac{\partial P}{\partial t}=\frac{\partial}{\partial \sigma}
\left(\frac{V'(\sigma)}{3H}\,P\right)+\frac12{\cal D} \frac{\partial^2P}{\partial \sigma^2}\,.
\]
In the case of fields with a non-negligible mass, as is the case of $\sigma$, the diffusion coefficient ${\cal D}$ features a mild scale dependence, which is studied in more detail in \cite{BS}. In the following, and for simplicity, we neglect this mild dependence and take ${\cal D}\simeq\frac{H^3}{4\pi^2}$. The first term in the righthand side accounts for the deterministic evolution of $P(\sigma,t)$, and depends on the scalar potential $V(\sigma)$. The second term accounts for the stochastic evolution of $P(\sigma,t)$, and its origin is the continuous outflow of perturbation modes $\delta\sigma_k$ crossing outside the horizon during inflation. This outflow of modes entails an imprint of structure in the classical field configuration on progressively smaller scales as inflation proceeds. The classical field $\sigma$ thus features correlations on comoving scales $k^{-1}$ and larger as long as $k^{-1}$ is outside the horizon. This reflects the fact that $P(\sigma,t)$ carries the probabilistic information concerning field correlations on all superhorizon scales at time $t$.

An aspect of $P(\sigma,t)$ of fundamental importance to the localization mechanism is its width. In the case of a massive field during slow-roll inflation, the width of the distribution is given by \cite{Starobinsky:1994bd}
\[\label{eq23}
\langle(\sigma-\bar\sigma)^2\rangle\simeq\frac{3H^4}{8\pi^2m_\sigma^2}
\left[1-\exp\left(-\frac{2m_\sigma^2t}{3H}\right)\right]\,,
\]
which can be obtained from Eq.~(\ref{eq12}) in the slow-roll limit for $c_\sigma\ll1$. From Eq.~(\ref{eq23}) it follows an important observation regarding the feasibility of the mechanism to generate the Cold Spot. Since the width of $P(\sigma,t)$ grows with time (as smaller scales exit the horizon), the largest deviation of $\sigma$ (away from $\bar\sigma$) is found in patches bearing field correlations on the smallest \emph{superhorizon} scales. As a result, to describe the statistics of the region where $\sigma\gtrsim\sigma_c$ (provided such region is non-vanishing), field correlations on the scale $k^{-1}$ become comparatively more important as $k$ grows. The importance of this to the mechanism relies on the fact that field correlations on small scales, even if abundant in the region $\sigma\gtrsim\sigma_c$ due to the larger expectation value, have a small chance to intersect the last scattering surface, and hence of arising in the CMB. On the other hand, regions where the field is correlated over larger distances, albeit comparatively less important (or even non-existing) to describe the region $\sigma\gtrsim\sigma_c$ due to the smaller field value, have nevertheless a greater chance to intersect the last scattering surface. The ensuing conclusion from this fact is that the $k$-patches that intersect the last scattering surface appear predominantly \emph{on a given scale}, while their presence on both larger and smaller scales is suppressed. The remaining of the section is dedicated to quantify this argument.

Returning to the probability density $P(\sigma,t)$, the usual procedure in the stochastic approach to inflation consists in using an approximately constant diffusion coefficient \mbox{${\cal D}=\frac{H^3}{4\pi^2}$} throughout the entire inflationary phase and then solving for Eq.~(\ref{eq15}). The density $P(\sigma,t)$ so computed carries the information on field correlations on all scales that are superhorizon by the end of inflation. For our purposes, however, we need to obtain the information on field correlations on a given comoving scale $k^{-1}$ only, hence we need to proceed differently. For our computation we consider a comoving scale $k^{-1}$ exiting the horizon at $t=t_k$. At that time, the solution to Eq.~(\ref{eq15}), $P(\sigma,t_k)$, carries the information on field correlation on scales $r\gtrsim k^{-1}$. Next, we need to evolve $P(\sigma,t_k)$ until the end of inflation. However, doing so by keeping the stochastic term in Eq.~(\ref{eq15}) leads to the imprint of structure on smaller scales. To obtain the information regarding field correlations on scales $r\gtrsim k^{-1}$ only, the imprint of structure must be shut down after $t=t_k$. This is equivalent to switching off the stochastic term in Eq.~(\ref{eq15}). The simplest alternative to do so is by introducing a scale-dependent cut-off in ${\cal D}$ to filter the appropriate modes. Thus, we solve for Eq.~(\ref{eq15}) using the diffusion coefficient
\[\label{eq16}
{\cal D}_k\equiv{\cal D}\,\theta(t_k-t)\,,
\]
where $\theta(t)$ is the step function. Taking as initial condition a distribution sharply peaked around $\sigma_*$ when the observable Universe exits the horizon at $t=t_*$, i.e. $P_k(\sigma,t_*)=\delta(\sigma-\sigma_*)$, we obtain a gaussian distribution $P_k(\sigma,t)$ whose mean and width $\Sigma^2(k,t)\equiv\langle(\sigma-\bar\sigma)^2\rangle$ at the end of inflation ($t=t_e$) are
\[\label{eq17}
\bar\sigma(t_e)=\sigma_*e^{-c_\sigma N_*/3}
\]
and
\[\label{eq18}
\Sigma^2(k,t_e)=\frac{3H^2}{8\pi^2 c_\sigma}\,e^{-\frac{2c_\sigma N_*}3}
\left[\left(\frac{k}{{\cal H}_*}\right)^{\frac{2c_\sigma}3}-1\right]\,,
\]
where ${\cal H}_*$ is the comoving Hubble scale at the onset of slow-roll. As shown later on, the scale dependence of $\Sigma^2(k,t_e)$ proves essential to the generation of $k$-patches in the appropriate range of scales.

The above parameters correspond to the solution of Eq.~(\ref{eq15}) in unbounded field space. However, given the trapping mechanism operating at $\sigma_c$, the Fokker-Planck equation must be supplemented with the so-called \textit{absorbing barrier} boundary condition: $P_k(\sigma_c,t)=0$ \cite{G4} (see also \cite{Lorenz:2010vf,Sanchez:2012tk} for applications to inflationary cosmology), and thus we should question the validity of Eqs.~(\ref{eq17}) and (\ref{eq18}) in the region $\sigma\geq\sigma_c$. The region $\sigma<\sigma_c$, on the other hand, becomes unphysical due to the presence of the barrier. For $\sigma\geq\sigma_c$, we must point out that as long as Eq.~(\ref{eq15}) (with ${\cal D}$ replaced by ${\cal D}_k$) is the usual Fokker-Planck equation, i.e. for $t\leq t_k$, its solution $P_k(\sigma,t)$ deviates significantly from a gaussian when it reaches the absorbing barrier, thus invalidating Eqs.~(\ref{eq17}) and (\ref{eq18}). On the other hand, the behavior of $P_k(\sigma,t)$ is very different if it reaches the barrier for $t>t_k$, thus implying that all the relevant structure is already imprinted in the field configuration when the distribution meets the barrier. Since Eq.~(\ref{eq15}) becomes a first order equation in this case, its solution $P_k(\sigma,t)$ is ``absorbed'' by the barrier without undergoing any distortion. This also implies a discontinuity in $P_k(\sigma,t)$ to satisfy the boundary condition. Since $P_k(\sigma,t)$ is not affected by the barrier for $\sigma>\sigma_c$, Eqs.~(\ref{eq17}) and (\ref{eq18}) hold exactly provided the scale $k^{-1}$ becomes superhorizon before the distribution hits the barrier. This turns out to be the situation for the range of scales probed in the CMB if $\sigma_*$ is sufficiently larger than $\sigma_c$ and $c_\sigma\lesssim{\cal O}(1)$. We address this question in more detail in \cite{BS}, where it will be shown that $\sigma_*$ can be consistently chosen so that Eqs.~(\ref{eq17}) and (\ref{eq18}) hold for all CMB scales in the entire range of the parameters $g$ and $c_\sigma$ allowed by the mechanism (see Eqs.~(\ref{eq21}) and (\ref{eq22})).

Using $P_k(\sigma,t)$ at the end of inflation (Eqs.~(\ref{eq17}) and (\ref{eq18})), the fraction of the inflated volume where $\sigma\gtrsim\sigma_c$ is correlated over comoving distances $r\gtrsim k^{-1}$ is
\[\label{eq3}
{\cal F}(k)=\int_{\sigma_c}^\infty \!\!P_k(\sigma,t_e)\,d\sigma=\frac12\left[1+\textrm{Erf}
\left(\frac{\bar\sigma(t_e)-\sigma_c}{\sqrt{2\Sigma^2(k,t_e)}}\right)\right]\,,
\]
which is scale-dependent owing to the evolution of the width during slow-roll. Clearly, ${\cal F}'(k)\,dk$ represents the fraction of the inflated volume with field correlations on scales in the interval $[k+dk,k]$. Using this fraction, the corresponding comoving volume in the observable Universe with field correlations in the above interval is $dV_k={\cal H}_*^{-3}\times{\cal F}'(k)\,dk$. Regarding the shape of $k$-patches, although the spatial region where a random gaussian field is above certain threshold (which corresponds to our definition of out-of-equilibrium patch after identifying the threshold with the crossover scale) can have a complicated structure (see for example \cite{random}), in \cite{Bardeen:1985tr} it was shown that the triaxial ellipsoid approximation is a valid description in the immediate neighborhood of the peak, and that high peaks tend to be more spherically symmetric than lower ones. In turn, nearly spherical shapes are only likely when very large thresholds (i.e. rarely occurring peaks) are considered. In the context of the Cold Spot, this fact is already appreciated in \cite{Afshordi:2010wn}. However, as we move away from the peak and encompass a larger spatial region, the triaxial ellipsoid approximation ceases to apply. In that case, the averaged shape of the region above the threshold (i.e. the out-of-equilibrium patch) tends to be spherically symmetric. Based on these results, it is then reasonable to conjecture that the roughly circular shape of the Cold Spot, inferred from its morphological analysis \cite{morpho}, originates as the intersection of an out-of-equilibrium patch with the last scattering surface. Also, the above results allow us to estimate that a $k$-patch (where $\sigma\gtrsim\sigma_c$ is correlated on the scale $k^{-1}$) occupies a typical volume $k^{-3}$, and hence that the typical number of $k$-patches is $dN_k=k^3dV_k$. Correspondingly, the number density $n(k)$ of $k$-patches per unit interval $dk$ \emph{in the observable Universe} is
\[\label{eq4}
n(k)\equiv dN_k/dk={\cal F}'(k)\left(\frac{k}{{\cal H}_*}\right)^3\,.
\]
Since the factor $(k/{\cal H}_*)^3$ dominates the scale dependence in the cases of interest, out-of-equilibrium regions are more abundant on small scales than on larger ones, as expected.

Concerning the field's spatial profile, after taking all the above into account it is reasonable to expect that, to some extent, the profile resembles a spherical top-hat. Such an expectation relies, on the one hand, on the different magnitude of the field between equilibrium and out-of-equilibrium regions (see Eq.~(\ref{eq9})) and, on the other hand, on the fact that the field value in out-of-equilibrium patches at the end of inflation can be easily seen to be of order $\sigma\sim\sigma_c$ \cite{BS}.

\subsection{Intersecting the last scattering surface}
Of course, the number density in Eq.~(\ref{eq4}) is not the desired, final result. If the anomalous CMB spots emerge as the intersection of a $k$-patch (for the appropriate $k$) with the last scattering surface, the number density in Eq.~(\ref{eq4}) must be multiplied times the corresponding probability of intersection. Regarding the location of $k$-patches, since these emerge as a result of the particle production undergone by $\sigma$ from its vacuum fluctuation, their spatial location is random. Invoking the separate Universe approach \cite{Wands:2000dp}, this implies that a $k$-patch has equal probability of emerging (provided it actually does) in any of the $(k/{\cal H}_*)^3$ independent patches of comoving size $k^{-3}$ contained in the volume ${\cal H}_*^{-3}$. Also, since the density in Eq.~(\ref{eq4}) corresponds to the number of $k$-patches in the observable Universe, the sought-after probability of intersection is subject to the condition that the $k$-patch emerges within the observable Universe. For the sake of simplicity, to estimate this probability we take the observable Universe as a box of side $2r_{\rm lss}$, where $r_{\rm lss}$ is the comoving radius of the last scattering surface, and $k$-patches to be spheres of comoving radius $r=k^{-1}/2$. Moreover, we assume that the size of the intersection between $k$-patches and the last scattering surface is of order $k^{-1}$ too. Under these assumptions, the probability that a $k$-patch intersects the last scattering surface, written as $P_{\rm lss}(r)=1-[P_{\rm in}(r)+P_{\rm out}(r)]$, where ``in'' (``out'') refers to the probability that the $k$-patch falls entirely inside (outside) the last scattering surface\footnote{We do not consider the case when two or more spheres intersect to give rise to an out-of-equilibrium patch of size larger than $r$ in the last scattering surface.}, can be easily computed.

To compute $P_{\rm in}(r)$, it suffices to realize that a $k$-patch falls entirely inside the last scattering surface whenever the center of the former, while randomly located in the observable Universe, lies within a sphere of radius $r_{\rm lss}-r$ centered at our location. For $r>r_{\rm lss}$, the probability $P_{\rm in}(r)$ is obviously zero, hence we multiply it times the step function \mbox{$\theta(r_{\rm lss}-r)$}. On the other hand, to compute $P_{\rm out}(r)$ it is enough to note that a $k$-patch falls entirely outside the last scattering surface when the distance between their centers is larger than $r_{\rm lss}+r$. The range of validity of this probability ($r\lesssim r_{\rm lss}$) can be extended to larger $r$ by defining $P_{\rm out}(r\geq r_{\rm lss})=0$. The probabilities $P_{\rm in,out}(r)$ determine the probability of intersection as a function of $r$. It is convenient, however, to rewrite the latter as a function of $(k/{\cal H}_*)$, similarly to Eq.~(\ref{eq4}). To do this, we note that since the scale of the observable Universe is currently entering the horizon, we have $k_{hor}={\cal H}_0={\cal H}_*$. Using also $r_{\rm lss}^{-1}\simeq{\cal H}_0/2$, we obtain \cite{BS}
\[\label{eq8}
P_{\rm lss}(k)=\frac{\pi}{2}\frac{{\cal H}_*}{k}\left[1+\frac1{12}\left(\frac{{\cal H}_*}{k}\right)^2\right]\,.
\]
Finally, multiplying Eqs.~(\ref{eq4}) and (\ref{eq8}) together we obtain the expected number density of $k$-patches per unit interval $dk$ in the last scattering surface:
\[\label{eq5}
{\cal N}(k)\equiv n(k)\,P_{{\rm lss}}(k)\simeq\frac{\pi}2\,{\cal F}'(k)\left(\frac{k}{{\cal H}_*}\right)^2\,.
\]

\begin{figure}[htbp]
\centering\epsfig{file=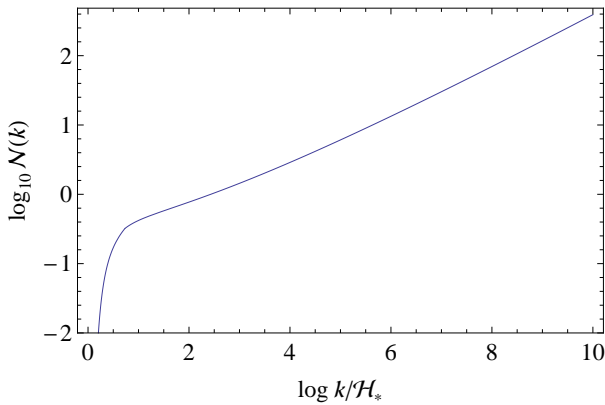,width=8cm}
  \caption{Number density ${\cal N}(k)$ of $k$-patches in the CMB. The plot illustrates the case \mbox{$c_\sigma=0.15$} and $g\sigma_*\simeq60H_*$.}\label{fig2}
\end{figure}

If the localization mechanism is to account for the Cold Spot by means of an overdensity in the last scattering surface, then we should expect to find \mbox{${\cal N}(k_{\rm cs})\sim{\cal O}(1)$}, where $k_{\rm cs}$ denotes the comoving scale of the Cold Spot. Note that the same condition must be imposed to account for anomalous hot spots through enhanced underdensities in the last scattering surface. To estimate $k_{\rm cs}/{\cal H}_*$, we use that the angle subtended by the Cold Spot on the last scattering surface is \mbox{$\vartheta_{\rm cs}\simeq 10^\circ$} \cite{Vielva:2010ng}, whereas $\vartheta_{\rm dec}\simeq1.7^\circ$ is the angle subtended by the horizon at decoupling. Assuming for simplicity a matter dominated Universe at present, we obtain \mbox{$\frac{k_{\rm cs}}{{\cal H}_*}\simeq\frac{\vartheta_{\rm dec}}{\vartheta_{\rm cs}}\,\Omega_m^{1/2}/(1+z_{\rm dec})^{1/2}\simeq3$}. We consider the case illustrated in Sec.~\ref{sec2b}, for which \mbox{$c_\sigma=0.15$} and \mbox{$g\sigma_*\simeq 60H_*$}, and plot the predicted number density ${\cal N}(k)$ in Fig.~\ref{fig2}. As shown in the plot, we find ${\cal N}(k)\sim{\cal O}(1)$ in the range $1\lesssim\log(k/{\cal H}_*)\lesssim3$, thus encompassing the Cold Spot scale.
Also, since ${\cal N}(k)$ grows with $k$, we should expect to find a larger number of out-of-equilibrium patches on smaller scales. Remarkably, this result is indeed consistent with the identification of several other anomalous spots in the CMB on scales slightly smaller than the corresponding to the Cold Spot \cite{Vielva:2007kt,Ade:2013nlj}. Regarding the magnitude of $\sigma_*$,  we can clarify now that this is so chosen since, for $c_\sigma=0.15$, the average field after $N_*=60$ $e$-foldings of slow-roll is $\bar\sigma(t_e)\simeq \sigma_c$. In turn, this implies that ${\cal F}(k)\sim{\cal O}(1)$, and hence the existence of $k$-patches in the range of scales of interest.

The plot in Fig.~\ref{fig2} also features a sharp fall on scales close to ${\cal H}_*^{-1}$. Such an abrupt fall owes to having assumed an absorbing barrier with zero thickness in field space, i.e. $P(\sigma_c,t)=0$, which is also responsible for the discontinuity of $P_k(\sigma,t)$ at the barrier previously noticed. This implies that for a range of scales very close to ${\cal H}_*^{-1}$, the distribution $P_k(\sigma,t_e)$, while sharply peaked around $\bar\sigma(t_e)$ (partly because of the $\delta$-like initial condition), is almost entirely either above the barrier (${\cal F}\simeq1$) or below \mbox{(${\cal F}\simeq0$)}. In any case, ${\cal F}'\simeq0$ for scales very close to ${\cal H}_*^{-1}$. On the other hand, an absorbing barrier with zero thickness is clearly an idealized situation, for it entails an instantaneous trapping of the field. In a more realistic situation, however, the trapping of the field occurs in the Hubble timescale. This is evident since the trapping of $\sigma$ relies on the production of the $\chi$ field from its vacuum fluctuation, which in turn takes place in the Hubble timescale. The implications of such a finite trapping time for the generation of $k$-patches is further discussed in \cite{BS}.

\subsection{A preferred scale in the CMB}\label{sec3b}
The growth of ${\cal N}(k)$ displayed in Fig.~\ref{fig2}, although consistent with the identification of several other anomalous spots in the CMB, also suggests that the effect of $k$-patches should be noticeable on smaller scales too. In particular, one might anticipate that if a $k$-patch leading to an enhanced overdensity in the last scattering surface is responsible for the Cold Spot, then a larger number of other $k$-patches should also have observational consequences on the scales affecting cosmological parameter fits ($\ell\geq50$). For example, one could expect to find the same level of non-Gaussianity implied by the Cold Spot, which would certainly result in $|f_{\rm NL}|\gg1$. However, the Planck data leave little room for primordial non-Gaussianity, establishing the consistency of the local, equilateral and orthogonal bispectrum amplitudes with zero at the $68$\% confidence level \cite{Ade:2013ydc}. Therefore, the non-detection of primordial non-Gaussianity suggests that the effect of $k$-patches on smaller scales must be negligible, which is seemingly incompatible with the generation of the Cold Spot by a $k$-patch. In the following, however, we show how the effect of $k$-patches on smaller scales indeed becomes negligible as $k$ increases.

\begin{figure}[htbp]
  \centering\epsfig{file=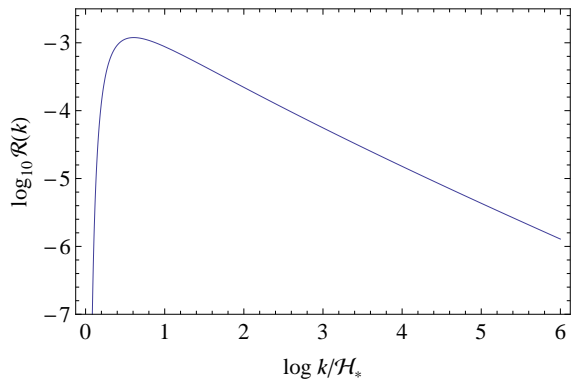,width=7.5cm}
  \caption{Relative number density ${\cal R}(k)$ of $k$-patches in the CMB, in the case $c_\sigma=0.15$ and $g\sigma_*\simeq 60H_*$. The predicted ratio peaks around the preferred scale $k_{\rm out}$.}\label{fig3}
\end{figure}

The key to understand why the effect of $k$-patches on scales smaller than the Cold Spot is imperceptible is very simple: the number of $k$-patches relative to the total number of patches of comoving size $k^{-1}$ in the last scattering surface, denoted by $n_{\rm lss}(k)$, becomes suppressed on smaller scales. To see this it suffices to note that the total number of regions of size $k^{-1}$ in the last scattering surface grows as $n_{\rm lss}(k)\propto k^2$. Using Eq.~(\ref{eq5}), we find the relative number density of $k$-patches per unit interval $dk$
\[\label{eq25}
{\cal R}(k)\equiv\frac{{\cal N}(k)}{n_{\rm lss}(k)}\simeq\frac{\pi}{2^7}\,{\cal F}'(k)\,.
\]
In Fig.~\ref{fig3} we depict the predicted ratio ${\cal R}(k)$ that follows from the number density ${\cal N}(k)$ plotted in Fig.~\ref{fig2}, i.e. for $c_\sigma=0.15$ and $g\sigma_*\simeq 60H_*$. The most salient feature of our result is that ${\cal R}(k)$ peaks around certain scale $k_{\rm out}$, becoming suppressed on both larger and smaller scales. Using Eqs.~(\ref{eq26}) and (\ref{eq17}) to (\ref{eq3}), the preferred scale $k_{\rm out}$ can be computed in terms of the model parameters solving for
\[\label{eq24}
{\cal F}''(k_{\rm out})=0\,.
\]
Moreover, using Eq.~(\ref{eq4}) we can also compute the ratio of isocurvature patches to adiabatic ones in the observable Universe, which is proportional to ${\cal F}'(k)$. Therefore, $k$-patches in the observable Universe also emerge preferentially on the scale $k_{\rm out}$. Since these give rise to overdensity regions, it would be worth investigating the consequences of the existence of $k$-patches at lower redshift, which is beyond the scope of this letter. Nevertheless, we note that the consequences of subhorizon bubbles at lower redshift have been examined in the context of the multifield inflation model of \cite{Afshordi:2010wn}.

According to the above discussion, $k$-patches are expected to have observational consequences in the CMB around the scale $k_{\rm out}$ only. This implies that on the range of scales relevant to cosmological parameter fits ($\ell\geq50$), the curvature perturbation imprinted in the CMB is almost entirely determined by the inflaton. Consequently, the simple fact that isocurvature $k$-patches are outnumbered by adiabatic ones on scales smaller than $k_{\rm out}^{-1}$, allows to account for the Cold Spot by means of a $k$-patch in the last scattering surface while, in principle, respecting the stringent bounds on non-Gaussianity imposed by Planck \cite{Ade:2013ydc}.

Before closing this section, we wish to point out that our localization mechanism for the generation of out-of-equilibrium patches could be applicable not only to scalars, but also to vector fields. In that case, cosmological vector fields obtain a spatial modulation similar to the one discussed in Sec.~\ref{sec2a}. Therefore, one might envisage a \emph{localized} version of the \emph{vector curvaton} scenario \cite{G6} to motivate a localized, direction-dependent contribution to the curvature perturbation, for example. This possibility is examined in more detail in a forthcoming publication \cite{BS}.

\section{Post-inflation evolution}\label{sec4}
In order to determine the curvature perturbation contributed by inhomogeneous reheating (see Eq.~(\ref{eq20})) we need to specify the evolution of $\sigma$ from the end of inflation, when the Universe becomes dominated by the matter-like oscillations of the inflaton, until the time of reheating. Since the inflaton oscillations provide a non-vanishing vacuum energy, in the following we envisage the persistence of the Hubble-induced correction to $m_\sigma^2$ until reheating \cite{Dine:1995uk}. In such case, the field equation for $\sigma$ is
\[
\ddot\sigma+3H\dot\sigma+c_\sigma H^2\sigma=0\,,
\]
where $H=\frac{2t}3$. The scaling of the growing mode (the dominant one after inflation) is $\sigma\propto a^{\gamma}$, where \mbox{$\gamma=-\frac34+\frac14\sqrt{9-16c_\sigma}$}. If \mbox{$16c_\sigma>9$}, $\gamma$ obtains an imaginary part, giving rise to field oscillations, and a real one that determines the scaling of the solution. This case describes the field dynamics in equilibrium patches, where $\sigma$ behaves as a heavy field due to its coupling to $\chi$. Taking the average over many oscillations, the field amplitude scales as $\sigma_{\rm eq}\propto a^{-3/4}$ in equilibrium patches. If $16c_\sigma<9$, the field does not oscillate about the origin, thus avoiding its non-perturbative decay\footnote{Note that the non-perturbative decay of $\sigma$ entails the disappearance of $k$-patches, thus rendering idle the mechanism here des\-cribed.}. This is the case of out-of-equilibrium patches. Keeping the first order in $c_\sigma$ in the expansion of $\gamma$, we find $\sigma_{\rm out}\propto a^{-2c_\sigma/3}$. This implies that the ratio $\sigma_{\rm out}/\sigma_{\rm eq}$ grows with time during the matter-dominated epoch. In turn, this keeps the feasibility of the mechanism since Eq.~(\ref{eq9}), and consequently Eq.~(\ref{eq7}), continue to hold. Moreover, it is straightforward to see that the energy density of $\sigma$ always remains subdominant; since the total energy density $\rho\propto H^2$ and $\rho_\sigma\simeq\frac12\,c_\sigma H^2\sigma^2$, it follows that $\rho_\sigma/\rho\propto a^{2\gamma}$, which always decreases with time.

An alternative, typical scenario is when $\sigma$ starts performing fast oscillations about the origin of its potential some time after inflation. Since the field oscillations may lead to the non-perturbative decay of $\sigma$, in \cite{BS} we examine this scenario assuming that $\sigma$ avoids such a decay.

\subsection{Constraints and feasibility of the mechanism}\label{sec4a}
In this section we constrain the model parameters using the condition that $\zeta_\sigma$ in Eq.~(\ref{eq20}) gives a sizable contribution to the total curvature perturbation. Since the ratio $\delta\sigma/\sigma$ remains constant, as pointed out before, to use Eq.~(\ref{eq20}) we must compute the ratio $(\delta\sigma/\sigma)_{\rm end}$.

Since the generation of $k$-patches in the appropriate scales demands that $\bar\sigma(t_e)\simeq\sigma_c$, at the end of inflation $\sigma$ has a typical value of order $\sigma_{\rm end}\sim g^{-1}H_*$ in out-of-equilibrium patches. On the other hand, the amplitude of the perturbation $\delta\sigma_k$ at the end of inflation can be computed using Eq.~(\ref{eq6}) in the slow-roll limit. Denoting by $N(k)$ the remaining number of $e$-foldings when $k$ exits the horizon, we find that \mbox{$(\delta\sigma)_{\rm end}\sim \frac{H_*}{2\pi}\,\exp[-N(k)\,c_\sigma/3]$}. Using these estimates, the contribution to the curvature perturbation becomes \cite{BS}
\[\label{eq19}
\zeta_\sigma\sim\frac{\alpha q r}{2\pi}\,g^{1-q}
\left(\frac{H_*}{M}\right)^q\left(\frac{T_{\rm rh}^2}{H_*m_P}\right)^{\frac{4qc_\sigma}9}e^{-N(k)c_\sigma/3}\,,
\]
which also applies to the decay rate in Eq.~(\ref{eq14}) after taking $q=2$, $r=1$ and substituting $M\to\sqrt{qr}\lambda^{-1}m_\phi$.

The parameters $g$ and $c_\sigma$ are constrained as follows. An appropriate initial value $\sigma_*$ must be large enough so that $g\sigma_*>H_*$, but sufficiently small for $\sigma$ to remain subdominant at the onset of slow-roll. Given $c_\sigma$, these two conditions are satisfied for $g\gg c_\sigma^{1/2}(H_*/m_P)$. The available space must be further constrained by the condition $\sigma_{\rm dec}<M$. From the assumed post-inflation evolution, this translates into $M>g^{-1}H_*(T_{\rm rh}/V_*^{1/4})^{8c_\sigma/9}$. Also imposing $g\lesssim1$, the allowed range of $c_\sigma$ and $g$ is determined by
\[\label{eq21}
0\leq c_\sigma<\frac3{N_*}\,\log\frac{\alpha q r}{2\pi\zeta_\sigma}
\]
and
\[\label{eq22}
\log\frac{2\pi \zeta_\sigma}{\alpha qr}+\frac{N_*}3\,c_\sigma<\log g\lesssim0\,,
\]
\begin{figure}[htbp]
\centering\epsfig{file=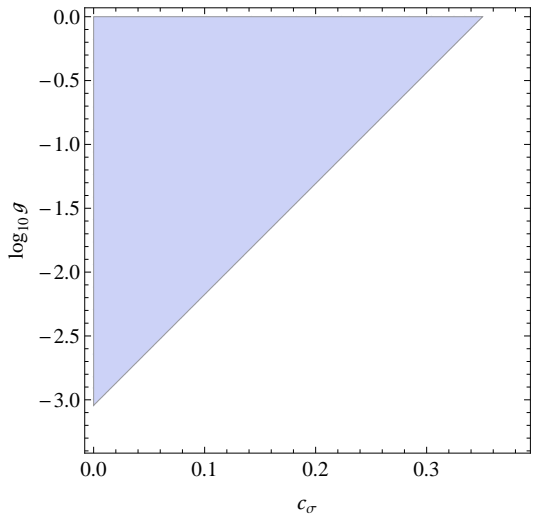,width=7.0cm}
  \caption{Available space for $g$ and $c_\sigma$, according to the constraints in Eqs.~(\ref{eq21}) and (\ref{eq22}), with $N_*=60$.}\label{fig4}
\end{figure}
which is stronger than $g\gg c_\sigma^{1/2}(H_*/m_P)$ for $q,r\sim{\cal O}(1)$ and $\zeta_\sigma\sim10^{-5}$. The range of parameters allowed by the above constraints is depicted in Fig.~\ref{fig4} for the particular case $q=2$, $r=1$, which makes the results applicable to the decay in Eq.~(\ref{eq14}) too. Fig.~\ref{fig5} depicts the range of $c_\sigma$ and $M$ (or $\sqrt{qr}\lambda^{-1}m_\phi$ equivalently) allowed by the same constraints and by the condition $\zeta_\sigma= 4.8\times10^{-5}$. The range shown in the figure corresponds to a reheating temperature in the interval $10^5 {\rm GeV}\leq T_{\rm rh}\leq10^9{\rm GeV}$. The plot demonstrates the existence of parameter space satisfying all the requirements, and therefore, the feasibility of the localized inhomogeneous reheating hypothesis to account for the Cold Spot for $c_\sigma\lesssim{\cal O}(1)$, according to theoretical expectations, through an enhanced overdensity in the last scattering surface. After replacing $M\to\sqrt{qr}\lambda^{-1}m_\phi$, the plot also demonstrates that anomalous hot spots can be accounted for through enhanced underdensities in the last scattering surface.
\begin{figure}[htbp]
\centering\epsfig{file=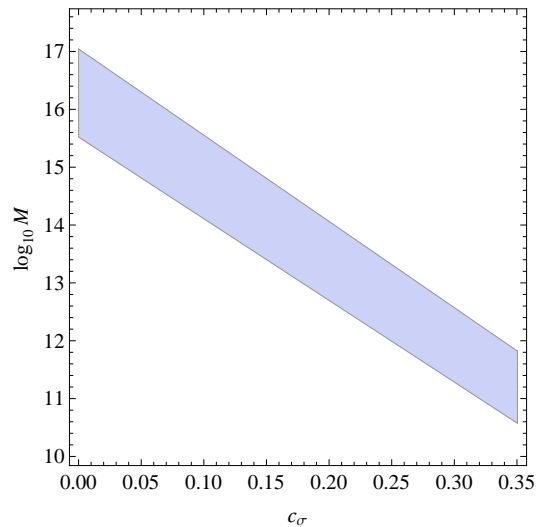,width=7cm}
  \caption{Available space for $c_\sigma$ and $M$ after imposing \mbox{$\zeta_\sigma=4.8\times10^{-5}$} and the constraints in Eqs.~(\ref{eq21}) and (\ref{eq22}).}\label{fig5}
\end{figure}

Finally, we remark that, owing to large mass correction considered in Eq.~(\ref{eq13}), the contribution $\zeta_\sigma$ to the curvature perturbation may feature a scale-dependent behavior. However, this may be difficult to detect since, as discussed in Sec.~\ref{sec3b}, the contributed $\zeta_\sigma$ has a chance to affect observations on scales close to $k_{\rm out}$ only (see Fig.~\ref{fig3}). In this sense, it is worth mentioning that when $k_{\rm out}$ matches the Cold Spot scale, the range of scales where observational consequences are expected, according to Fig.~\ref{fig3}, has an important overlapping with the corresponding to the multipoles $20\lesssim\ell\lesssim40$, where WMAP and Planck report an unusual shape of the spectrum \cite{G5,Ade:2013zuv}. It is thus tempting to suggest that such an unusual shape can be related to the emergence of $k$-patches in the CMB.

\section{Conclusions}\label{sec5}
In this letter we have studied a system of two interacting, massive fields, $\sigma$ and $\chi$, coupled to each other through the interaction $g^2\sigma^2\chi^2$. We allow both fields to receive corrections to their masses of order the Hubble scale, but such that this correction, by itself, does not keep the fields from being produced from their vacuum fluctuation during inflation. In this simple setting, a mechanism arises whereby the expectation value of $\sigma$ features a spatial modulation by the end of inflation. Such a modulation arises during the last phase of slow-roll (here assumed to last 60 $e$-foldings) only if $\sigma$ begins this phase with an expectation value sufficiently above the Hubble scale so that $\chi$ becomes a heavy field. This condition, however, poses a difficulty for the mechanism. The reason is that fields with masses in the Hubble scale typically have expectation values of order $H$ (see Eq.~(\ref{eq23})), which then turns the required initial condition into an unnatural one. Nevertheless, we demonstrate that if the Universe undergoes a non-slow-roll phase of inflation, the typical value of $\sigma$ indeed grows above the Hubble scale (see Eq.~(\ref{eq12})).

In order to convert the spatial modulation of $\sigma$ into a curvature perturbation, we develop a localized version of the inhomogeneous reheating hypothesis. In this, the contribution to the curvature perturbation $\zeta_\sigma$ in out-of-equilibrium patches manages to affect the one imprinted by the inflaton. On the contrary, in equilibrium regions where $\sigma$ is trapped, the contributed curvature perturbation $\zeta_\sigma$ is negligible (see Eq.~(\ref{eq7})). Consequently, the distribution of out-of-equilibrium patches attending to their typical size is an aspect of utmost importance to the localization mechanism. Using simple assumptions, we estimate the number density of $k$-patches (spatial regions of size $k^{-1}$ where $\sigma$ has an out-of-equilibrium value) at the end of inflation, Eq.~(\ref{eq4}). The strong scale dependence of $n(k)$ is an expected result, since scales with larger $k$ exit the horizon later, when the typical value of $\sigma$ is larger. After estimating the probability of intersection between a $k$-patch and the last scattering surface, Eq.~(\ref{eq8}), we obtain an estimate for the number density of $k$-patches in the CMB, Eq.~(\ref{eq5}).

In Fig.~\ref{fig2}, we show that a reasonable choice of parameters suffices to obtain ${\cal N}(k)\sim{\cal O}(1)$ around the Cold Spot scale at the end of inflation. The case illustrated corresponds to $c_\sigma=0.15$, for which the initial condition $g\sigma_*\simeq 60H_*$, necessary to have $k$-patches of the appropriate size, can be justified after around 20 \mbox{e-foldings} of inflation (for $g$ in the range $0.1\leq g\leq1$) with $\epsilon=0.3$. Also, the Hubble scale at the onset of the non-slow-roll phase varies in the range $2.4\times10^{17}\,{\rm GeV}\geq H_0\geq 3.4\times10^{16}\,{\rm GeV}$, thus setting the beginning of inflation close to the Planck scale, as expected on theoretical grounds. Since the initial condition for the emergence of the Cold Spot is set by the background dynamics previous to slow-roll inflation, our approach suggests to use the Cold Spot as a tool to probe the earliest phase of inflation, before our observable Universe exits the horizon. Moreover, our approach leads us to interpret the Cold Spot as the signal of an \emph{out-of-equilibrium remnant} of an isocurvature field, which managed to survive thanks to the conjunction of two facts. First, the large expectation value obtained by the isocurvature field $\sigma$ during the non-slow-roll phase, and second, the relatively short phase of slow-roll inflation that follows and during which fields with masses on the Hubble scale relax to their equilibrium values.

Regarding the scale dependence of ${\cal N}(k)$, remarkably, this is consistent with the generation of several other anomalous spots, already identified in the CMB on a slightly smaller scale. However, in spite of this appealing feature, the scale dependence of ${\cal N}(k)$ raises the concern that the effect of $k$-patches should be noticeable on scales smaller than the Cold Spot, which would possibly ruin the adiabaticity, scale-invariance and Gaussianity of the observed curvature perturbation. Nevertheless, as demonstrated in Fig.~\ref{fig3} (see also Eq.~(\ref{eq25})), ${\cal N}(k)$ grows slower than the total number of patches of size $k^{-1}$ in the CMB. And the same conclusion applies to the number of $k$-patches in the observable Universe [c.f. Eqs.~(\ref{eq4}) and (\ref{eq5})]. As a result, the effect of $k$-patches on smaller scales becomes suppressed simply because they are outnumbered by the patches of the same size where the inflaton imprints its adiabatic, nearly scale-invariant, gaussian curvature perturbation. Interestingly, Fig.~\ref{fig3} evidences the existence of a scale $k_{\rm out}$, determined by particle physics and inflationary parameters (see Eqs.~(\ref{eq26}), (\ref{eq17})-(\ref{eq3}) and (\ref{eq24})), around which the existence of isocurvature $k$-patches can have observational consequences.

Apart from imposing ${\cal N}(k)\sim{\cal O}(1)$ around the Cold Spot scale, the model parameters must be constrained so that the curvature perturbation $\zeta_\sigma$ (see Eq. (\ref{eq19})) contributed by inhomogeneous reheating is of order $10^{-5}$ . Fig.~\ref{fig4} shows the range of $g$ and $c_\sigma$ compatible with the mechanism (see Eqs.~(\ref{eq21}) and (\ref{eq22})), whereas Fig.~\ref{fig5} shows the allowed range for $M$ (see Eq.~(\ref{eq2})), or $\sqrt{qr}\lambda^{-1}m_\phi$ equivalently (see Eq.~(\ref{eq14})), after imposing $\zeta_\sigma=4.8\times10^{-5}$ and the constraints in Eqs.~(\ref{eq21}) and (\ref{eq22}). Also, to build Fig.~\ref{fig5} we allow the reheating temperature to take on values in the interval $10^5\,{\rm GeV}\leq T_{\rm rh}\leq10^9\,{\rm GeV}$. The plot \emph{demonstrates} that the localized inhomogeneous reheating hypothesis can indeed account for the Cold Spot, provided the latter is due to an enhanced overdensity in the last scattering surface, using reasonable values of the particle physics and inflationary parameters. Since the Cold Spot constitutes the most significant deviation from Gaussianity, the scenario here presented provides a physical mechanism to generate a \emph{localized} non-gaussian signal in the CMB. Moreover, Fig.~\ref{fig5} also demonstrates that anomalous hot spots in the CMB can be accounted for through enhanced underdensities in the last scattering surface.

To conclude, the setting presented in this letter provides a physical mechanism to generate the Cold Spot anomaly in the CMB through the enhancement of an overdensity in the last scattering surface. We argue that if the Cold Spot is caused by a recently discovered supervoid, the localized inhomogeneous reheating, along with the appropriate decay rate (see Eq.~(\ref{eq2})), may help resolve the question of its rarity, as it results in enhanced underdensities. Furthermore, the localization mechanism predicts the generation of several other anomalous spots on a slightly smaller scale, which, on the other hand, have already been identified in both the WMAP and Planck data. Therefore, having a number of interesting avenues to investigate, this work offers new ground to explore the origin of the recently confirmed CMB anomalies.

\acknowledgements
The author wishes to thank M. Bastero-Gil, K. Dimopoulos and M. Cruz for comments and discussions. The author is supported by COLCIENCIAS grant No. 110656399958.

\end{document}